\documentstyle[11pt]{article}   
\begin{document}   
\thispagestyle{empty}   
\rightline{UOSTP-01103}  
\rightline{KIAS-P01016}   
\rightline{{\tt hep-th/0103148}}   
   
\

\def\tr{{\rm tr}\,} \newcommand{\beq}{\begin{equation}}   
\newcommand{\eeq}{\end{equation}} \newcommand{\beqn}{\begin{eqnarray}}   
\newcommand{\eeqn}{\end{eqnarray}} \newcommand{\bde}{{\bf e}}   
\newcommand{\balpha}{{\mbox{\boldmath $\alpha$}}}   
\newcommand{\bsalpha}{{\mbox{\boldmath $\scriptstyle\alpha$}}}   
\newcommand{\betabf}{{\mbox{\boldmath $\beta$}}}   
\newcommand{\bgamma}{{\mbox{\boldmath $\gamma$}}}   
\newcommand{\bbeta}{{\mbox{\boldmath $\scriptstyle\beta$}}}   
\newcommand{\lambdabf}{{\mbox{\boldmath $\lambda$}}}   
\newcommand{\bphi}{{\mbox{\boldmath $\phi$}}}   
\newcommand{\bslambda}{{\mbox{\boldmath $\scriptstyle\lambda$}}}   
\newcommand{\ggg}{{\boldmath \gamma}} \newcommand{\ddd}{{\boldmath   
\delta}} \newcommand{\mmm}{{\boldmath \mu}}   
\newcommand{\nnn}{{\boldmath \nu}}   
\newcommand{\diag}{{\rm diag}}   
\newcommand{\bra}{\langle}   
\newcommand{\ket}{\rangle}   
\newcommand{\sn}{{\rm sn}}   
\newcommand{\cn}{{\rm cn}}   
\newcommand{\dn}{{\rm dn}}   
\newcommand{\tA}{{\tilde{A}}}   
\newcommand{\tphi}{{\tilde\phi}}   
\newcommand{\bpartial}{{\bar\partial}}   
\newcommand{\br}{{{\bf r}}}   
\newcommand{\bx}{{{\bf x}}}   
\newcommand{\bk}{{{\bf k}}}   
\newcommand{\bq}{{{\bf q}}}   
\newcommand{\bQ}{{{\bf Q}}}   
\newcommand{\bp}{{{\bf p}}}   
\newcommand{\bP}{{{\bf P}}}   
\newcommand{\thet}{{{\theta}}}   
\newcommand{\tauu}{{{\tau}}}   
\renewcommand{\thefootnote}{\fnsymbol{footnote}}   
\   
   
\vskip 0cm   
\centerline{  
\Large   
\bf Noncommutative  Supersymmetric Tubes   
}   
 
\vskip .2cm   
 
\vskip 1.2cm 
\centerline{  
Dongsu Bak$^a$ 
  and Kimyeong Lee$^{b}$
} 
\vskip 10mm  
\centerline{ \it $^a$ Physics Department,  
University of Seoul, Seoul 130-743, Korea}  
\vskip 3mm  
\centerline{ \it $^b$ School of Physics, Korea Institute for Advanced  
Study}  
\centerline{ \it 207-43, Cheongryangri, Dongdaemun, Seoul  
130-012, Korea  
}   
\vskip 0.3cm 
\centerline{\tt 
( dsbak@mach.uos.ac.kr, 
klee@kias.re.kr )}

\vskip 1.2cm    
\begin{quote}   
{ We investigate supersymmetric tubular configurations in the matrix
theory.  We construct a host of BPS configurations of eight
supersymmetries.  They can be regarded as cylindrical D2 branes
carrying nonvanishing angular momentum.  For the simplest tube, the
world volume can be described as noncommutative tube and the world
volume dynamics can be identified as a noncommutative gauge
theory. Among the BPS configurations, some describe excitations on the
tube and others describe many parallel tubes of different size and
center. }
\end{quote}   
   

\newpage   
\section{Introduction}  

It is now well known how to describe the half BPS D-branes from the
view point of matrix theory\cite{Banks,Bigatti,Seiberg1}.  The D-brane
configurations arise as a flat hypersurface formed by constituent
D-particles.  There was also some study of a spherical D2
configurations and its variations~\cite{Kabat}.  When external R-R
fields are turned on, the constituent D-particles respond to the
external field like a dielectric material in an external
electromagnetic field\cite{Emparan,Myers}.  For example, a spherical D2-brane
configuration of finite radius occurs and the geometry involved here
is so called a fuzzy sphere.  The bending of the surface, in this
case, is caused by the external R-R fields.  It is found recently a
tube configuration from the Born-Infeld theory description, which is
self sustained by its worldvolume gauge field\cite{Mateos}.  The
configuration is a quarter BPS state unlike the other D-brane
configurations and also carrying a nonvanishing angular momentum
produced by the worldvolume gauge field.

In this paper we like to realize the tube configuration from the view
point of the matrix theory. To this end, we first derive a set of BPS
equation and find solutions describing tubes. The worldvolume geometry
is defined by an algebra describing a noncommutative cylinder.  The
configuration involves the worldvolume magnetic field as well as
electric field, which leads to a nonvanishing angular momentum.  The
direction of worldvolume electric field is found to agree with the
supertube of the Born-Infeld theory\cite{Mateos}.  However the precise
match of the charges, the strength of the electric field, and so on
are not clear because highly nonlinear terms play a role in the
Born-Infeld description.

Since both electric and magnetic fields are 
present in the worldvolume, one may naively expect that
its dynamics is described by a spacetime 
noncommutative gauge theory. But it turns out 
that the worldvolume dynamics  is described by
a gauge theory involving spacelike noncommutativity
only. However the presence of electric field makes
the gauge theory differ from the conventional 
noncommutative 
Yang-Mills theory on a cylinder. We develop here 
also the $*$-product realization of the algebra and operators
on the noncommutative supersymmetric tube. 
This makes the geometrical 
interpretation  clear. 

We construct noncommutative soliton configurations describing multiple
D0's out of the noncommutative tube.  Here it is not difficult to
identify the moduli parameters involved with the
solitons. Interestingly, the moduli dimensions are the number of D0's
multiplied by nine--the number of spatial dimensions. Hence the D0
solitons seem to easily fly off the tube at least classically.  Unlike
ordinary unstable D0-D2 system\cite{Aganagic}, the configurations are
BPS saturated preserving again eight supersymmetries.  The tube
solution may be generalized to 
coincident tubes. The worldvolume dynamics here become $U(p)$
noncommutative gauge theory\cite{Bak3}.  We also find many tubes of
different sizes and centers.

In Section 2, we present the tube solution of the matrix 
model.  The low energy description of 
the worldvolume gauge theory is investigated in Section 3.
In Section 4, we construct  the solitonic solutions 
describing D-particles out of the tube. In addition, we discuss 
more general configuration of many tubes.
Last section comprises conclusions and comments.

\section{Supersymmetric Tube Solutions}   

To construct the tube  
configurations, we shall begin with the matrix  
model  Lagrangian 
\begin{equation}   
L={1\over 2 R} \tr \left( (D_0 X_I)^2 
+{R^2\over 2\, l^6_{11}} [X_I,X_J]^2+ {\rm fermionic\ part} 
\right)  
\label{lag}   
\end{equation}   
where $I,J=1,2,\cdots 9$ and $R$ is the  
radius of tenth  
spatial direction.  Here $l_{11}$ is the eleven dimensional 
Planck length, which we will set to unity in the following 
discussions. The scale $R$ will be omitted below for 
simplicity and we shall recover it whenever necessary. 
As is well known, this Lagrangian can be thought of  
describing $N$ D-particles if one takes all the dynamical 
variable as $N\times N$ matrices.  
 
Let us first describe relevant BPS equations  
we like to solve. For this, we shall turn on only first three  
components of the matrices $X_I$. Then the Gauss law reads 
\begin{equation}   
[X,D_0 X]+[Y,D_0 Y]+[Z,D_0 Z]=0\,. 
\label{gausslaw}   
\end{equation}   
Using the Gauss  constraint, the bosonic part of the
Hamiltonian can be written as 
\begin{equation}   
H={1\over 2 } \tr \left( (D_0 X\pm i[Z,X])^2+(D_0 Y\pm i[Z,Y])^2+(D_0 Z)^2 
+ [X,Y]^2+ 2 C_J 
\right)\  \ge\   \tr C_J 
\label{bound}   
\end{equation}   
where  $\tr C_J$  is the central charge  defined by  
\begin{equation}   
\tr C_J= \pm{i\over 2}\tr \sum^3_{i=1}\, [X_i\,, Z (D_0 X_i)+(D_0 X_i) Z]\,.  
\label{centralcharge}   
\end{equation}    
The saturation of the BPS bound occurs if the BPS equations 
\begin{eqnarray}   
&&  [X,Y]=0,\ \ \ \ \ \ \ D_0 Z=0\,,\nonumber\\ 
&&\ D_0 X\pm i[Z,X]=0, \ \ \ D_0 Y\pm i[Z,Y]=0  
\label{bpsequations}   
\end{eqnarray}   
hold together with the Gauss law constraint in (\ref{gausslaw}). 
On the choice of gauge $A_0={R\over \, l^3_{11}}\, Z$, the 
BPS equations of the upper sign  imply that all the fields are 
static. Hence in this gauge, the system of equations 
 reduce to 
\begin{eqnarray}   
[X,Y]=0,\ \ \ \  [X,[X,Z]]+[Y,[Y,Z]]=0\,, 
\label{BPS}   
\end{eqnarray}  
where the latter comes from the Gauss law constraint.
Before providing the  
representation of the algebra, let us count the remaining supersymmetries  
of the state specified by the nontrivial representation of the the  
algebra.  
There are two sixteen supercharges of the matrix model. The 16  
components of the kinematical supersymmetry is broken spontaneously 
by the presence of the longitudinal momentum. 
The remaining supersymmetric variation of the  fermionic 
coordinates $\psi$ is given by  
\begin{eqnarray}   
  \delta \psi =  -D_0 X (\Gamma_{01} + \Gamma_{13})\epsilon 
-D_0 Y  (\Gamma_{02} + \Gamma_{23})\epsilon\,. 
\label{susy}   
\end{eqnarray}   
Setting this to zero, one finds that the solutions 
preserve eight of remaining sixteen supersymmetries. 
Hence in total the configuration preserves a quarter  
of the 32 supersymmetries of the matrix model.

Among the solutions of the BPS equations, we are particularly  
interested in the solutions defined by the following algebra, 
\begin{eqnarray}   
  [z,x]=il y, \ \ [y,z]=i l x\,, \ \   [x,y]=0,  
\label{bpssolutions}   
\end{eqnarray}   
with $X_i=x_i$. 
This is the algebra defining two dimensional Euclidean group. The length  
scale $l$ is the noncommutativity parameter.  
 
The algebra in (\ref{bpssolutions}) is realized as follows. 
Let us introduce variables $x_\pm$ by 
\begin{eqnarray}   
x_\pm= x\pm i y\,. 
\label{xpm}   
\end{eqnarray}   
The algebra is then rewritten as\footnote{The same algebra
is considered in Ref.\cite{Chaichian} in the context of
spacetime noncommutativity.} 
\begin{eqnarray}   
  [z,x_\pm]=\pm l  x_\pm, \ \ [x_-,x_+]= 0 \,,  
\label{algebra}   
\end{eqnarray}   
so it is clear that
$x_-x_+ =x^2+y^2\equiv\rho^2$ is a  
Casimir operator. 
We are interested in the following irreducible 
representation 
of the algebra, 
\begin{eqnarray}   
x_+|n\ket=\rho |n+1\ket\,,\ \ \  
z|n\ket=l n |n\ket \, .
\label{represenation}   
\end{eqnarray}   
We use $|n\ket$ ($n\in {\bf Z})$ to be the basis to represent infinite
dimensional matrices.

As $\rho^2$ is Casimir operator and can be regarded as a number, we
can represent the $x_\pm$ with the angular variable as follows
\begin{equation}
x_\pm = \rho e^{\pm i \theta}
\end{equation}
with periodic hermitian operator $\theta$. Then $e^{\pm i\theta}|n\ket
= |n\pm 1\rangle$ and $[z,e^{\pm i \theta}] = \pm l e^{\pm i\theta} $.
It is obvious that our BPS configuration describes a noncommutative
tube of radius $\rho$ in three dimensions.  The coordinates
$(\theta,z)$ of this tube would be noncommutative.

Any well-defined operator   can be presented as 
\begin{eqnarray}   
f(z,\theta) =\sum^\infty_{n=-\infty}\int^{\pi\over l }_{-{\pi\over l}} 
{dk\over 2\pi} \tilde{f}_n (k) 
e^{i n \theta+ik z}\,. 
\label{function}   
\end{eqnarray}  
The range of $k$ is determined by the fact that the $z$ operator has
discrete eigenvalues. Also any operator can be represented as 
$ f= \sum^\infty_{n,m=-\infty} f_{nm}|n\ket\bra m|$ in the matrix
theory. Two representations are related by
\begin{eqnarray}   
f_{nm}=\int^{\pi\over l}_{-{\pi\over l}} 
{dk\over 2\pi} \tilde{f}_{n-m} (k) 
e^{ {i l\over 2} (n+m)k}  
\,. 
\label{component}   
\end{eqnarray}  
For the operation relation, we also get  $|0\rangle\langle 0| = l\int
{dk\over 2\pi} e^{ikz}$.

The multiplication of operators on the  noncommutative tube is well
defined. Instead, we can define the $*$-product of ordinary functions
on the corresponding commutative  tube of the same radius. In the
fourier representation of ordinary functions, their $*$-product should
leads to the  fourier 
representation which we would get as the product of operators. Thus,
the $*$ product of two ordinary functions $g$ and $h$ would be
\begin{eqnarray}   
g* h= \left[e^{{il\over 2}(\partial_\theta \partial_{z'}- 
\partial_z \partial_{\theta'} 
)}  g(\theta, z) h(\theta',z')\right]_{\theta=\theta', z=z'}\,.  
\end{eqnarray}  
In addition, the spatial integration $\int d\theta dz$ on the tube
corresponds to $2\pi l \tr $.  Since above $*$-product implies that
$\theta * z- z* \theta =il$, the minimal area is in a rough sense
given by $\Delta= (2\pi l)\times \rho$.  Since the circumference of
the tube is $2\pi \rho$, one may regard the noncommutativity scale $l$
as a kind of minimal distance in the $z$ direction. Indeed the
discreteness of the spectrum of $z$ is consistent with this
observation.  Moreover, $1/\Delta$ corresponds to the area density of
the the constituent D0-branes.
Namely,  the total number of D0-branes is $ N=\tr I= {1\over
\Delta}\int dz  d\theta \rho$. 
Hence the D0 brane density per unit length in the z-direction  
is $1/l$.    
 
Now let us reconsider  the  preliminary discussion of the physical  
implication of the solution. 
First of all, 
the central charge can be reexpressed as 
\begin{equation}   
\tr C_J= l^2 \tr x_- x_+  = l^2 \rho^2 \tr I\,,  
\end{equation}    
where we have used the algebra in (\ref{algebra}). 
The central charge density per unit length in z-direction is 
$l \rho^2 R /l^6_{11}$. 
We compare this with the angular momentum along the $z$ axis, 
\begin{equation}   
\tr J=\tr(X D_0 Y-Y D_0 X )= -l \rho^2 \tr I\,.  
\end{equation}    
Thus $\tr C_J= -l\tr J$ and the system carries a nonvanishing  
angular momentum density.  
As we seen above, the configuration describes a tubular 
configuration whose coordinates may be identified as 
$\theta$ and $z$ in the commutative limit. It is a cylindrical  
object embedded in a flat 9-dimensional space. 
Hence we shall call 
the configuration as noncommutative supersymmetric tube.

{}From the view of the commutative Born-Infeld action, there exists a
nonzero magnetic field along the $\rho$ direction, which is
responsible for the noncommutativity.  In addition, there exists an
electric field on the world volume along the $z$ direction.  Since $
D_0 X= l y$, $ D_0 Y=-lx$, it might appear that the electric field is
applied to the $\theta$ direction. But this identification is not
quite right because the open string metric describing the worldvolume
dynamics is twisted due to the presence of background $B_{\theta z}$
field\cite{Seiberg}.  In short, the configuration discussed above
corresponds to the supertube found recently in the Born-Infeld
description\cite{Mateos}.
 
\section{Worldvolume Gauge Theory}
 
In this section, we shall describe the low energy dynamics of the
worldvolume gauge theory by taking the solution (\ref{bpssolutions})
as a background configuration. For this end, let us first consider the
field equation governing fluctuation of the transverse scalar.
Turning on just one component, $\phi=X_4$, the equation reads
\begin{equation}   
[\partial_t-iz,[\partial_t-iz,\phi]]+ 
[x_i,[x_i,\phi]]=0 \,,  
\label{scalareq}  
\end{equation}  
where, recovering the $l_{11}$ and $R$,  
$x_i$ is replaced by $R x_i/l^3_{11}=x_i/l^2_s$ with  
the string scale $l_s$. First, let us write the  
equation in the commutative limit. To this end, 
we note 
\begin{eqnarray}   
&&[z,\cdot]= -i l \partial_\theta\,, 
\ \ \  
[x,\cdot]= -ily \partial_z +O(l^2)\,, 
 \nonumber\\  
&&[y,\cdot]= il x \partial_z +O(l^2)\,. 
\label{approximation}   
\end{eqnarray}   
Using this relation, the above equation can be written as 
\begin{equation}   
(\partial_t-l\partial_\theta)^2\phi- 
l^2 \rho^2 \left({1\over  \rho^2} \partial_\theta^2 +\partial_z^2\right) 
\phi 
+O(l^3) 
=0 \,. 
\end{equation}
This is the equation of motion produced by the Lagrangian
density
\begin{equation}   
{\cal L}_\phi={1\over 2}\left(\partial_t\phi\right)^2
-l\partial_t\phi \,\partial_\theta\phi
-{l^2 \rho^2\over 2}\left(\partial_z\phi\right)^2 
+O(l^4)\,. 
\label{philag}   
\end{equation}  
The system is a field theoretic analog of electrically charged
 particles moving in a constant magnetic field.  One can find the
 canonical momentum density $p_\phi = \dot{\phi} -l \partial_\theta
 \phi$ and its Hamiltonian density $H_\phi = (p_\phi+l\partial_\theta
 \phi)^2/2 +\rho^2 (\partial_z \phi)^2/2$, which is positive definite.

As is well known in the case of constant magnetic field, the effect of
electric field can be canceled by going to the rotating frame defined
by
\begin{equation}   
z'=z\,,\ \ \theta'= \theta+ l t 
\,, \ \ \ t'=t  
\label{rotation}   
\end{equation}    
The equation then becomes a free field equation on  
a cylinder: 
\begin{equation}   
\partial_{t'}^2\phi(\theta',z')- 
l^2 \rho^2 \left({1\over  \rho^2}  
\partial_{\theta'}^2 +\partial_{z'}^2\right) 
\phi (\theta',z') 
+O(l^3) 
=0 \,.  
\end{equation}  
 
In fact the scalar equation in (\ref{scalareq}) can be solved
generically without difficulties.  To this end, let us use the
operator representation $\phi=\sum_n \int {dk\over 2 \pi} 
\tilde{\phi}_n(k,t)
e^{in(\theta +lt )+ikz} $, where we introduce the time dependent phase
to cancel the effect of the background electric field. The scalar
equation for each fourier component becomes trivial and its  most
general solution  is then 
\begin{eqnarray}   
\tilde{\phi}_n(k,t)= f_n(k)  e^{\pm i\left[l^2 n^2+ 4\rho^2  
\sin^2 ({lk\over 2})\right]^{1\over 2}\, t} 
\, 
\end{eqnarray}    
with  arbitrary $f_n(k)$. 
 
We now turn to the description of the world volume gauge theory. The
$U(\infty)$ gauge symmetry of the matrix theory becomes a local gauge
symmetry of the world volume theory with local gauge transformation 
$U(t,\theta,z)$. Two of $\Delta X_I$ fluctuations around the 
supersymmetric tube solution would act as the gauge field and the
rest of them as the scalar field. However, the world volume action is
more complicated than the naive noncommutative Yang-Mills theory
because the tube is imbedded in the three dimensional space and also
there is a nonzero background momentum. The resulting Lagrangian is not
that illuminating. Thus, we focus on the small fluctuation around on
the tube configuration. To derive the Lagrangian governing the
dynamics of these fluctuations in the leading order, we begin with the
Lagrangian in (\ref{lag}) and turn on again only $X_i$ for simplicity.
We now introduce gauge fields $a_0$, $a_\theta$, $a_z$, and $a_\rho$
by
\begin{eqnarray}   
&&X=x-{l\over 2}(y a_z+ a_z\, y) + {l\over 2}(x a_\rho+ 
a_\rho \, x) + O(l^3)\nonumber\\ 
&&Y=y+{l\over 2}(x a_z+ a_z\, x) + {l\over 2}(y a_\rho+ 
a_\rho \, y) + O(l^3)\nonumber\\ 
&&Z=z-l a_\theta\,, \ \ \ A_0=z+l a_0  
\label{app}   
\end{eqnarray}   
In evaluating the Lagrangian, we shall count orders of fields by  
the noncommutativity scale $l$ by 
\begin{eqnarray}   
&& a_0\,,\ a_\theta\,,\   a_z\,,\  
a_\rho\   \sim\  O(l^0)\nonumber\\ 
&& [a_\theta,a_z]\,,\  
 [a_\theta,a_\rho]\,,\  [a_\rho,a_\rho]\ \sim \ O(l^0) \nonumber\\ 
&& \partial_t\ \sim \  O(l)\,. 
\end{eqnarray}   
Especially the second line  might appear   
unconventional but is consistent. 
Using (\ref{approximation}), 
 the commutators become 
\begin{eqnarray}   
&& [X,Y]= -i l^2 \rho^2 \nabla_z a_\rho + O(l^3)\nonumber\\ 
&& [X,Z]+il y= \ l^2 y  
 \Bigl([\nabla_z,\nabla_\theta]-ia_\rho\Bigr)+ il^2 x 
\Bigl(\nabla_\theta a_\rho-a_z\Bigr)+ O(l^3) \nonumber\\ 
&& [Y,Z]-il x= -l^2 x  
\Bigl( [\nabla_z,\nabla_\theta]-ia_\rho\Bigr)+ 
il^2 y \Bigl(\nabla_\theta a_\rho-a_z\Bigr)+ O(l^3)\,, 
\end{eqnarray}     
where $\nabla_\theta\equiv \partial_\theta -i[a_\theta,\cdot]$,
$\nabla_\rho\equiv \partial_\rho -i[a_\rho,\cdot]$
and $\nabla_z\equiv \partial_z -i[a_z,\cdot]$ . 
One recognizes that the worldvolume coordinates 
are 
twisted from the original matrix  
coordinate. For example, the commutator $[X,Y]$ is not related 
to the z-directional magnetic field. 
Inserting these to the Lagrangian directly, one would get a 
 Lagrangian where terms of  $O(l^3)$ are present and 
higher order terms in (\ref{app}) appear in general.
Instead, we add  appropriate total derivative terms
first so that the leading order contribution starts with terms of 
$O(l^4)$ and the higher order terms in (\ref{app}) appear only
in the next leading order contributions.    
Following this procedure, one gets
\begin{eqnarray}   
L&=&{\,l^2\over 2 } \tr \left(   
(\dot{a}_\theta -l\nabla_\theta \tilde{a}_0)^2 + 
 \rho^2 (\dot{a}_z -il\nabla_z \tilde{a}_0
)^2 
+ \rho^2 (\dot{a}_\rho -il [\tilde{a}_0, a_\rho])^2
-l^2\rho^4 (\nabla_z a_\rho)^2  
\right.\nonumber\\ 
&&\left.  
-2i l\rho^2 (\dot{a}_\theta -l\nabla_\theta \tilde{a}_0)
[\nabla_z,\nabla_\theta]
-2 l \rho^2 (\dot{a}_\rho -il [\tilde{a}_0, a_\rho])
\nabla_\theta a_\rho +2 l^2 \rho^2{\cal L}_{\rm CS}  
+O(l^5)\right)\,,  
\end{eqnarray}
where $\tilde{a}_0=a_0 + a_\theta$ and 
\begin{eqnarray}   
\tr {\cal L}_{CS}=-\partial_t a_z a_\rho + a_z\partial_t a_\rho
- 2 l\tilde{a}_0 \left(\partial_z a_\rho -i[a_z,a_\rho]
\right)\,.
\label{cs}   
\end{eqnarray}
This is  the standard form of the Chern-Simons Lagrangian
with $\partial_\rho=0$.
As done in the scalar case, we now introduce the rotating 
coordinate system in (\ref{rotation}). Then the above Lagrangian 
becomes  
\begin{eqnarray}   
L&=&{\, l^2\over 2 } \tr \left(   
-[\nabla_{t'}, \nabla_{\theta'}]^2   
-\rho^2  [\nabla_{t'}, \nabla_{z'}]^2  
+ l^2 \rho^2 [\nabla_{z'},\nabla_{\theta'}]^2
+\rho^2 (\nabla_{t'} a_{\rho'})^2
\right.\nonumber\\ 
&&\ \ \ \left. 
-l^2\rho^2 (\nabla_{\theta'} a_\rho)^2 
-l^2\rho^4 (\nabla_{z'} a_\rho)^2  
+2 l^2\rho^2 {\cal L}_{\rm CS}
+O(l^5)\right)\,,  
\end{eqnarray}   
where $a_{\theta'}=a_\theta$,  $a_{z'}=a_z$, $a_{\rho'}=a_\rho$
and $\nabla_{0'}\equiv \partial_{t'} -i[a_{0'},\cdot]$ 
with $a_{0'}=a_0$. 
The theory contains a Chern-Simons term in addition to the
conventional U(1) noncommutative gauge theory on a cylinder with the
 rotating coordinate $\theta'=\theta+l t$.  The Chern Simons part
 plays a role in finding D0 solutions in the next section.  Namely the
 solutions of D0 excitations preserve the same supersymmetries of the
 tube background. If the worldvolume theory were just conventional
 noncommutative Yang-Mills theories, such higher supersymmetric
 solutions could not exist.

In the next section, we shall describe the nature of  
solitonic configurations arising from the tube. The  
configurations correspond to adding or subtracting 
multiple D0's from the tube. To the given order of approximation, 
these solutions can be found  from the above action,  
but we shall rather seek the solutions from the original  
 equations of motion.

\section{Tube-D0 Systems and Multiple Tubes} 

In this section, we shall first consider other static solutions that
approach asymptotically the tube configurations. For this, one may try
to solve the equations of motion in the background of the D-tube
configuration concentrating on the solutions describing localized
profiles. This is the approach taken in finding the noncommutative
solitons in various two dimensional
models\cite{Gopakumar,Bak,Aganagic,Harvey}, which may produce BPS
configurations in some special cases\cite{Bak,Witten}.  Instead, we
will solve directly the BPS equations while preserving the tubular
boundary conditions., that is, we solve the equations in (\ref{BPS}).
In fact it is simple to construct such localized profiles on the
D-tube background.  First let us introduce a shift operator defined by
\begin{eqnarray}   
S_m = \sum^\infty_{n=0} |n+m\ket \bra n| 
+ \sum_{n=-\infty}^{-1} |n\ket \bra n|\,. 
\end{eqnarray}     
It satisfies the relations 
\begin{eqnarray}   
S_m S_m^\dagger= I-P_m\,,\ \ \  
S^\dagger_m S_m =I\,, 
\end{eqnarray}      
where the projection operator $P_m$ is defined by $
P_m=\sum^{m-1}_{a=0} |a\ket \bra a| $. 
Then the solutions describing $m$ D0-branes are given by 
\begin{eqnarray}   
X_i=S_m \, x_i S_m^\dagger\,. 
\end{eqnarray}    
Unlike the  case of the noncommutative Yang-Mills theory
describing a planar D2-brane, 
the solutions we constructed here are 
BPS saturated states of eight supersymmetries.

If one computes the central charges corresponding to the D0
configurations, one naively gets $\tr C_J= l^2 \rho^2 \tr
(I-P_m)$. The difference with the value for the background indicates
that the energy has been lowered.  This kind of problem
is not new.  The noncommutative Yang-Mills theory on D2-branes for
example may be related to the matrix theory\cite{Gopakumar}.  Then the
matrix theory computation of energy for the D2-D0 system produce the
result like $\tr (I-P_m)$.  There, from the view point of the
noncommutative Yang-Mills theory, the noncommutative D0 solutions are
well localized configurations certainly carrying finite energy excited
above the vacuum of the noncommutative Yang-Mills theory that is well
defined at least classically.  Moreover, the fluctuation spectra
around these solutions perfectly agree with the worldsheet conformal
theory analysis of the superstrings.

We expect that a similar resolution may exist for our present problem.
However, as our supersymmetric vacuum (\ref{bpssolutions}) and the
 excitations discussed above have the same eight 
supersymmetries, there should be some crucial modification in the
world volume field theory. In ordinary case, we expect the excited BPS
configuration to have lower number of supersymmetry than the
vacuum. This difference needs a further analysis.

The more general solutions including the moduli parameters
are given by
\begin{equation} 
 X_i=S_m \, x_i S_m^\dagger +
\sum^{m-1}_{a=0}\lambda_i^{a}|a\ket\bra a| \, , \,\,\,\, 
 X_s= \sum^{m-1}_{a=0}\varphi_s^{a}|a\ket\bra a|\,, 
\end{equation}
with the index $s$ referring to the transverse scalar $X_4$
to $X_9$. 
The moduli are describing the positions of D0 branes
in the 9-dimensional space. The appearance of moduli further
support the view point that the configurations
are describing not holes in tubes but extra  D0-branes
that may even fly off the tube.

Next we like to mention briefly another type of solutions which
describe many coincident  tubes.  The solution is given by
\begin{eqnarray}   
X+iY=    \rho \sum^\infty_{n=-\infty} \sum^{p-1}_{a=0} 
|(n+1)p +a\ket\bra np+a|\,,\ \ \  
Z= l \sum^\infty_{n=-\infty}\sum^{p-1}_{a=0}
n |np +a\ket\bra np+a|\,,   
\label{concentric}
\end{eqnarray}   
where $p$ is a positive integer characterizing the solutions.  This
background makes the worldvolume theory being a $U(p)$ noncommutative
gauge theory. The $U(p)$ basis can be constructed by writing
$|np+a\ket \bra mp +b|=|n\ket'\bra m|' T_{ab}$.  Here $|n\ket'$ is
interpreted as a new basis for the space while $T_{ab}$ generates
$U(p)$ algebra.  For the further details, see Ref.\cite{Bak3} where
the $U(p)$ vacuum solutions of the noncommutative gauge theory is
constructed in the U(1) noncommutative gauge theory on a plane.

This type of solutions in (\ref{concentric}) 
can be generalized  further
to describe many parallel tubes whose centers are located in arbitrary
positions on the $(X,Y)$ plane. These solutions are 
\begin{eqnarray}  
X+iY &=& \sum_{a=0}^{p-1} \rho_a \sum_{n=-\infty}^{\infty}
 |(n+1)p+ a\rangle\langle np+a|
+  \sum^{p-1}_{a=0} \xi_a
 \sum^\infty_{n=-\infty}|np +a\ket\bra np+a|\, , \nonumber  \\
Z &=& \sum_{a=0}^{p-1} l_a \sum_{n=-\infty}^\infty n\,|n p+a\ket \bra
np+a| 
\, ,   
\end{eqnarray}  
where $\rho_a$ is for the radius of each tube, $l_a$ is for the
noncommutative parameter of each tube, and $\xi_a$ is for the position
of the center of each tube. Again they are BPS saturated
configurations preserving eight supersymmetries.  The identification
of the worldvolume gauge theory for these general configurations is
not clear at this point, as the noncommutative parameters for tubes
can be different from each other. Of course, we can add the position
along the other dimensions and also excitations on each tube to get a
further generalization of the above solutions.

\section{Conclusion}  

In this note, we found BPS configurations describing tubes from the
matrix model. The low energy description of the worldvolume gauge
dynamics is also identified.  We find soliton solutions describing
many D0's on the tube.  In addition, we found BPS solutions describing
many coincident tubes, whose worldvolume dynamics is $U(p)$
noncommutative gauge theory. There are further solutions of many
parallel tubes of different size and center.

As we found additional supersymmetric configurations besides the
supersymmetric tubes found in Ref.~\cite{Mateos}. It would be
interesting to see whether its analogue exists in the Born-Infeld
theory.  The set of BPS equations in (\ref{BPS}) in a static gauge
appears quite simple. Also it would be interesting to see if there
exist other category of BPS solutions besides those found  here.
 
The relation between supersymmetry and the world volume dynamics of
the noncommutative supersymmetric tubes needs a further consideration
as discussed in the previous section. Additional understanding of the
tube dynamics may be obtained from the approach of worldsheet
conformal field theory of superstrings\cite{Chen}. 
Namely as done in
Ref.\cite{Aganagic} for the D2-D0, the conformal field theory
description may provide detailed dynamical information on the tube-D0
systems. The results then may be compared to the fluctuation analysis
around the tube-D0 configurations.

\noindent{\large\bf Acknowledgment} This work is supported in part by
KOSEF 1998 Interdisciplinary Research Grant 98-07-02-07-01-5 (DB,KL)
and by UOS Academic Research Grant (DB).
 


\end{document}